\documentstyle[preprint,floats,pra,eqsecnum,aps,psfig]{revtex}
\tightenlines
\begin{document} \draft

\title{Does Lorentz Boost Destroy Coherence?\footnote{Presented at
the Workshop on Fundamental Problems in Quantum Theory, Baltimore,
Maryland, U.S.A. (August 1997); to be published in the Proceedings.}}

\author{Y. S. Kim}
\address{Department of Physics, University of Maryland, College Park,
Maryland 20742, U.S.A.}

\maketitle

\begin{abstract}
It is shown that the time-energy uncertainty relation can be
combined into the position-momentum uncertainty relation
covariantly in the quark model of hadrons.  This leads to a
Lorentz-invariant form of the uncertainty relations.  This
model explains that the quark model and the parton model are two
different manifestations of the same covariant model.  In particular,
this covariant model explains why the coherent amplitudes in the
quark model become incoherent, after a Lorentz boost,
in the parton model.  It is shown that this lack of coherence
is consistent with the present form of quantum mechanics.

\end{abstract}

\section{Introduction}\label{intro}
At this conference, there are a number of papers dealing with the
time variable.  The reason is very simple.  We cannot do physics
without this variable.  Its role is well defined in Newton's equation
in classical mechanics.  However, the time variable becomes
complicated when we move to quantum mechanics and to relativity.
In classical mechanics, we use the horizontal axis for the time
variable and the vertical axis for the position or momentum.
However, at this conference whose purpose is to address fundamental
questions of quantum mechanics, we have seen a number of papers with
space-time diagrams in which the vertical axis is used for the time
variable.

Quantum mechanics is not the only place where we use the vertical
axis for the time variable.  We have been doing this in special
relativity since it was formulated by Einstein in 1905.  Does this
mean that quantum mechanics is becoming closer to special relativity?
The purpose of this report to say YES to this question.  In quantum
mechanics, where position and momentum are constrained by the
uncertainty relation, we are still arguing about the whether
there is an uncertainty relation between the time and the energy
variables.  As for the position-energy uncertainty relation,
Heisenberg's uncertainty relation is stated by the canonical
commutation relations between position and momentum operators.
For the time variable, we are not allowed to write down a
commutation relation between the time and energy variables because
there are no time-like excitations in the real world.  However,
the time-energy uncertainty clearly is clearly observed in the
world.  How to accommodate this space-time asymmetry into the
space-time symmetric relativistic world has been one of the most
outstanding problems in physics since 1927~\cite{dir27}.

These days, it is a routine laboratory procedure to accelerate the
proton to the energy one thousand times higher than its rest mass.
While we regard the proton as a bound state of quarks when it is at
rest or slow~\cite{gell64}, the question arises whether we can use the
existing rules of quantum mechanics to understand the proton whose
speed is very close to that of light.  In 1969~\cite{fey69}, Feynman
observed that the high-energy proton is a collection of free
particles with a wide-spread momentum distribution.  They appear to
be incoherent when they interact with external signals.  If then the
high-energy proton is a Lorentz-boosted proton at from its rest frame,
the coherence observed in the quark model is destroyed by Lorentz boost.

Does the Lorentz boost destroy the coherence?  We shall study this
question in connection with the task of combining bound-state quantum
mechanics with special relativity.  In Sec.~\ref{covar}, we spell out
this problem in terms of the known principles of quantum mechanics
and special relativity.  In Sec.~\ref{covham}, the problem is
formulated in terms of the covariant harmonic oscillators.  In
Sec.~\ref{parton}, it is explained how the peculiarities in the
Feynman's parton picture arise from the covariance of quantum mechanics.
In Sec.~\ref{cohere}, we explain that the lack of coherence in the
parton picture is consistent with the present form of quantum mechanics
and therefore that the Lorentz boost does not destroy the coherence.

\section{Covariance and Quantum Mechanics}\label{covar}
To physicists, Einstein's $E = mc^{2}$ means $E = \sqrt{p^{2} +
m^{2}}$.  This Lorentz-covariant quantity leads to
$E = p^{2} /2m$ and $E = cp$ in the limits of low and high speeds
respectively.
In addition, relativistic particles have internal space-time
degrees of freedom.  For a massive particle, there is always a
Lorentz frame in which the particle is at rest.  In this Lorentz
frame, the particle has the three-dimensional rotational degrees
of freedom.  The dynamical quantity associated with this degree of
freedom is the intrinsic angular momentum called the spin.  For a
massless particle, however, there are no Lorentz frames where the
particle is at rest.  The particle in this case has the angular
momentum either parallel or antiparallel to the momentum.  It
does not have the rotational symmetry the massive particle has in
its rest frame.  In addition, the massless particle has a gauge
degree of freedom.

It is still one of the most fundamental questions in physics to
ask whether the gauge degree of freedom can be regarded as a
space-time transformation.  This issue has a stormy history, but
it has now been firmly established that the transverse rotational
degrees of freedom become a gauge degree of freedom in the
infinite-momentum/zero-mass limit.  This feature is illustrated
the second row of Table~I.

\begin{table}

\caption{Covariance of Relativistic Particles.  In addition to the
four-momentum, the particle has internal space-time symmetries.  The
spin and gauge symmetries are tabulated in the second row.  The
particle can also have a space-time extension, which manifests itself
as the quark model and the parton model.}

\vspace{4mm}

\begin{tabular}{cccc}

{}&{}&{}&{}\\
{} & Massive, Slow \hspace*{1mm} & COVARIANCE \hspace*{1mm}&
Massless, Fast \\[4mm]\hline
{}&{}&{}&{}\\
Energy- & {}  & Einstein's & {} \\
Momentum & $E = p^{2}/2m$ & $ E = [p^{2} + m^{2}]^{1/2}$ & $E = cp$
\\[4mm]\hline
{}&{}&{}&{}\\
Internal & $S_{3}$ & {}  &  $S_{3}$ \\[-0.5mm]
Space-time &{} & Wigner's  & {} \\ [-0.5mm]
Symmetry & $S_{1}, S_{2}$ & Little Group & Gauge Trans. \\[4mm]\hline
{}&{}&{}&{}\\
Relativistic & {} & Covariant Model  &   {} \\[-0.5mm]
Extended & Quark Model & of & Parton Model \\ [-0.5mm]
Particles & {} & Bound States & {} \\[2mm]
\end{tabular}
\end{table}

In order to arrive at the conclusion of the second row, we have to
study the little groups of the Poincar\'e group.
The Poincar\'e group is the group of inhomogeneous Lorentz
transformations, namely Lorentz transformations preceded or followed
by space-time translations.  In order to study this group, we have to
understand first the group of Lorentz transformations, the group of
translations, and how these two groups are combined to form the
Poincar\'e group.  The Poincar\'e group is a semi-direct product of
the Lorentz and translation groups.  The two Casimir operators of
this group correspond to the (mass)$^{2}$ and (spin)$^{2}$ of a given
particle.  Indeed, the particle mass and its spin magnitude are
Lorentz-invariant quantities.

The next question is how to construct the representations of the
Lorentz group which are relevant to physics.  For this purpose,
Wigner in 1939 studied the subgroups of the Lorentz group whose
transformations leave the four-momentum of a given free
particle invariant~\cite{wig39}.  The maximal subgroup of the Lorentz
group which leaves the four-momentum invariant is called the little
group.  This little group governs the internal space-time symmetries of
relativistic particles.  Wigner shows in his paper that the internal
space-time symmetries of massive and massless particles are dictated
by the $O(3)$-like and $E(2)$-like little groups respectively.

The group of Lorentz transformations consists of three boosts and
three rotations.  The rotations therefore constitute a subgroup of
the Lorentz group.  If a massive particle is at rest, its four-momentum
is invariant under rotations.  Thus the little group for a massive
particle at rest is the three-dimensional rotation group.  Then what is
affected by the rotation?  The answer to this question is very simple.
The particle in general has its spin.  The spin orientation is going
to be affected by the rotation!

There are no Lorentz frames where a massless particle is at rest.
If the massless particle moves along the $z$ direction, rotations
around the its momentum leave the four-momentum invariant.  In addition,
there are two generators of the Lorentz group which leave the momentum
invariant.  If we take the commutation relations of these three
generators of the little group, they are exactly like those for the
two-dimensional Euclidean group which we call $E(2)$.   The group $E(2)$
consist of translations and rotations on a flat surface.  It is not
difficult to associate the rotational degree of freedom with the
helicity of the massless particle.  But the generators of the
translation-like transformations have a stormy history.  They generate
gauge transformations when applied to the four-potential~\cite{janner71}.

If the rest-particle is boosted along the $z$ direction, it will pick
up a non-zero momentum component.  The generators of the $O(3)$ group
will then be boosted.  The boost will take the form of conjugation by
the boost operator.  This boost will not change the Lie algebra of the
rotation group, and the boosted little group will still leave the
boosted four-momentum invariant.  We call this the $O(3)$-like little
group.  The question then is whether the $O(3)$-like little group
becomes the $E(2)$-like little group in the high-speed limit.

The question of Lorentz-boosted Poincar\'e group in the
infinite-momentum limit was addressed first by Bacry and Chang in
1968 in connection with scattering problems~\cite{bacryc68}.  Since
the $O(3)$-like little group is a subgroup of the Poincar\'e group,
Bacry and Chang in effect obtained the $E(2)$-like little group as
the infinite-momentum limit of a subgroup of the Lorentz group.  This
high-speed contraction of the little group was later
modeled after the Inonu-Wigner contraction of the $O(3)$ to $E(2)$
as a flat-surface approximation of a spherical surface with a large
radius~\cite{inonu53}.  It was found later that the transverse
rotation generators become contracted to the generators of the
translation-like transformations~\cite{ferrara82}.  Indeed, the
rotations around the transverse directions become contracted to gauge
transformations in the limit of infinite momentum and/or zero
mass~\cite{hks83}.

Next, let us summarize quantum mechanics.  Quantum field theory
has been quite successful in terms of perturbation techniques in
quantum electrodynamics.  However, this formalism is basically based
on the S matrix for scattering problems and useful only for physically
processes where free a set of particles becomes another set of free
particles after interaction.  Quantum field theory does not address
the question of localized probability distributions and their
covariance under Lorentz transformations.

The Schr\"odinger quantum mechanics of the hydrogen atom deals with
localized probability distribution.  Indeed, the localization condition
leads to the discrete energy spectrum.  Here, the uncertainty relation
is stated in terms of the spatial separation between the proton and
the electron.  If we believe in Lorentz covariance, there must also
be the time separation between the two constituent particles.  This
does not manifests itself in nonrelativistic quantum mechanics, but
it exists.  The time interval seems to be an important issue at this
conference.

Indeed, we have to add a time dimension to spatial coordinates before
getting into the relativistic world.  As we noted in Sec.~\ref{intro},
there are many papers in this conference with space-time diagrams with
the time coordinate as the vertical axis.  This is precisely what we do
in relativity.  When we make a Lorentz boost along the $z$ direction,
the transformation is written as
\begin{equation}\label{boostm}
\pmatrix{z' \cr t'} = \pmatrix{\cosh \eta & \sinh \eta \cr
\sinh \eta & \cosh \eta } \pmatrix{z \cr t} ,
\end{equation}
This formula is well known, but it is not yet widely known that this
is a squeeze transformation.  In order to see this point, let us use
the light-cone variables defined as~\cite{dir49}
\begin{equation}
u = (z + t)/\sqrt{2} , \qquad v = (z - t)/\sqrt{2} .
\end{equation}
Then the boost transformation of Eq.(\ref{boostm}) takes the form
\begin{equation}\label{lorensq}
u' = e^{\eta } u , \qquad v' = e^{-\eta } v ,
\end{equation}
where $\eta $ is the boost parameter and is $\tanh ^{-1}(v/c)$.
The $u$ variable becomes expanded while the $v$ variable becomes
contracted.  This is the squeeze mechanism discussed extensively in
the literature~\cite{kn73,knp91}.  The Lorentz boost is a squeeze
transformation.  By now, the word ``squeeze'' is quite familiar to us
from the squeezed states of light.  Thus, the first step in making
quantum mechanics covariant is to work out carefully a space-time
picture of non-relativistic quantum mechanics in one Lorentz frame.
The next step is to squeeze the space-time diagram.

Let us be more specific.  Before 1964~\cite{gell64}, the hydrogen
atom was used for illustrating bound states.  These days, we use
hadrons which are bound states of quarks.  Let us use the simplest
hadron consisting of two quarks bound together an attractive force.
For the probability distribution, we can use the Gaussian form for
spatial separation between the quarks.  This spatial coordinates are
quantized, and the position and momentum variables are q-numbers.

There is also the time-energy uncertainty relation applicable to
the time separation between the quarks.  Unlike Heisenberg's
uncertainty relation applicable to position and momentum, the
time and energy separation variables are c-numbers, and we are not
allowed to write down the commutation relation.  On the other hand,
the c-number time energy uncertainty relation allows to write down a
time distribution function without excitations.  If we use Gaussian
forms for both space and time distributions, we can start with the
expression
\begin{equation}
\exp{\left\{-{1 \over 2}\left(z^{2} + t^{2}\right)\right\}} ,
\end{equation}
where the $z$ and $t$ are the space and time separations respectively.
The present form of quantum mechanics allows the excitations along
the $z$ direction, but there are no excitations along the $t$ direction.
Yet, we can start from a circular space-time distribution given by the
above expression.  We can then boost the distribution by squeezing it.
We are therefore able to start from a hadron at rest, and boost it
to the infinite-momentum frame.

For the third row in Table~I, we propose to solve the following problem
in high-energy physics and foundations of quantum mechanics.  The quark
model works well when hadrons are at rest or move slowly.  However,
when they move with speed close to that of light, they appear as a
collection of infinite-number of partons~\cite{fey69}.  As we stated
above, we need a set of wave functions which can be Lorentz-boosted.
How can we then construct such a set?  In constructing wave functions
for any purpose in quantum mechanics, the standard procedure is to try
first harmonic oscillator wave functions.  In studying the Lorentz boost,
the standard language is the Lorentz group.  Thus the first step to
construct covariant wave functions is to work out representations of
the Lorentz group using harmonic oscillators~\cite{dir45,yuka53,knp86}.

\section{Covariant Harmonic Oscillators}\label{covham}

If we construct a representation of the Lorentz group using normalizable
harmonic oscillator wave functions, the result is the covariant harmonic
oscillator formalism~\cite{knp86}.  The formalism constitutes a
representation of Wigner's $O(3)$-like little group for a massive
particle with internal space-time structure.  This oscillator formalism
has been shown to be effective in explaining the basic phenomenological
features of relativistic extended hadrons observed in high-energy
laboratories.  In particular, the formalism shows that the quark model
and Feynman's parton picture are two different manifestations of one
covariant entity~\cite{knp86,kim89}.  The essential feature of the
covariant harmonic oscillator formalism is that Lorentz boosts are
squeeze transformations~\cite{kn73,knp91}.  In the light-cone coordinate
system, the boost transformation expands one coordinate while contracting
the other so as to preserve the product of these two coordinate remains
constant.  We shall show that the parton picture emerges from this
squeeze effect.

Let us consider a bound state of two particles.  For convenience, we
shall call the bound state the hadron, and call its constituents quarks.
Then there is a Bohr-like radius measuring the space-like separation
between the quarks.  There is also a time-like separation between the
quarks, and this variable becomes mixed with the longitudinal spatial
separation as the hadron moves with a relativistic speed.  There are
no quantum excitations along the time-like direction.  On the other
hand, there is the time-energy uncertainty relation which allows
quantum transitions.  It is possible to accommodate these aspect within
the framework of the present form of quantum mechanics.  The uncertainty
relation between the time and energy variables is the c-number
relation~\cite{dir27},
which does not allow excitations along the time-like coordinate.  We
shall see that the covariant harmonic oscillator formalism accommodates
this narrow window in the present form of quantum mechanics.

For a hadron consisting of two quarks, we can consider their space-time
positions $x_{a}$ and $x_{b}$, and use the variables
\begin{equation}
X = (x_{a} + x_{b})/2 , \qquad x = (x_{a} - x_{b})/2\sqrt{2} .
\end{equation}
The four-vector $X$ specifies where the hadron is located in space and
time, while the variable $x$ measures the space-time separation between
the quarks.  In the convention of Feynman {\it et al.} \cite{fkr71},
the internal motion of the quarks bound by a harmonic oscillator
potential of unit strength can be described by the Lorentz-invariant
equation
\begin{equation}\label{osceq}
{1\over 2}\left\{x^{2}_{\mu} -
{\partial ^{2} \over \partial x_{\mu }^{2}}
\right\} \psi (x)= \lambda \psi (x) .
\end{equation}
It is now possible to construct a representation of the Poincar\'e group
from the solutions of the above differential equation~\cite{knp86}.

The coordinate $X$ is associated with the overall hadronic
four-momentum, and the space-time separation variable $x$ dictates
the internal space-time symmetry or the $O(3)$-like little group.  Thus,
we should construct the representation of the little group from the
solutions of the differential equation in Eq.(\ref{osceq}).  If the
hadron is at rest, we can separate the $t$ variable from the equation.
For this variable we can assign the ground-state wave function to
accommodate the c-number time-energy uncertainty relation~\cite{dir27}.
For the three space-like variables, we can solve the oscillator
equation in the spherical coordinate system with usual orbital and
radial excitations.  This will indeed constitute a representation of
the $O(3)$-like little group for each value of the mass.  The solution
should take the form
\begin{equation}
\psi (x,y,z,t) = \psi (x,y,z) \left({1\over \pi }\right)^{1/4}
\exp \left(-t^{2}/2 \right) ,
\end{equation}
where $\psi(x,y,z)$ is the wave function for the three-dimensional
oscillator with appropriate angular momentum quantum numbers.  Indeed,
the above wave function constitutes a representation of Wigner's
$O(3)$-like little group for a massive particle \cite{knp86}.

Since the three-dimensional oscillator differential equation is
separable in both spherical and Cartesian coordinate systems,
$\psi(x,y,z)$ consists of Hermite polynomials of $x, y$, and $z$.
If the Lorentz boost is made along the $z$ direction, the $x$ and $y$
coordinates are not affected, and can be temporarily dropped from the wave
function.  The wave function of interest can be written as
\begin{equation}
\psi^{n}(z,t) = \pmatrix{{1\over \pi }}^{1/4}\exp \pmatrix{-t^{2}/2}
\psi_{n}(z) ,
\end{equation}
with
\begin{equation}
\psi ^{n}(z) = \left({1 \over \pi n!2^{n}} \right)^{1/2} H_{n}(z)
\exp (-z^{2}/2) ,
\end{equation}
where $\psi ^{n}(z)$ is for the $n$-th excited oscillator state.
The full wave function $\psi ^{n}(z,t)$ is
\begin{equation}\label{2.6}
\psi ^{n}_{0}(z,t) = \left({1\over \pi n! 2^{n}}\right)^{1/2} H_{n}(z)
\exp \left\{-{1\over 2}\left(z^{2} + t^{2} \right) \right\} .
\end{equation}
The subscript $0$ means that the wave function is for the hadron at rest.
The above expression is not Lorentz-invariant, and its localization
undergoes a Lorentz squeeze as the hadron moves along the $z$
direction~\cite{knp86}.

The wave function of Eq.(\ref{2.6}) can be written as
\begin{equation}\label{10}
\psi ^{n}_{o}(z,t) = \psi ^{n}_{0}(z,t)
= \left({1 \over \pi n!2^{n}} \right)^{1/2} H_{n}\left((u + v)/\sqrt{2}
\right) \exp \left\{-{1\over 2} (u^{2} + v^{2}) \right\} .
\end{equation}
If the system is boosted, the wave function becomes
\begin{equation}\label{11}
\psi ^{n}_{\eta }(z,t) = \left({1 \over \pi n!2^{n}} \right)^{1/2}
H_{n} \left((e^{-\eta }u + e^{\eta }v)/\sqrt{2} \right)
\times \exp \left\{-{1\over 2}\left(e^{-2\eta }u^{2} +
e^{2\eta }v^{2}\right)\right\} .
\end{equation}

In both Eqs. (\ref{10}) and (\ref{11}), the localization property of
the wave function in the $u v$ plane is determined by the Gaussian
factor, and it is sufficient to study the ground state only for the
essential feature of the boundary condition.  The wave functions in
Eq.(\ref{10}) and Eq.(\ref{11}) then respectively become
\begin{equation}\label{13}
\psi _{0}(z,t) = \left({1 \over \pi} \right)^{1/2}
\exp \left\{-{1\over 2} (u^{2} + v^{2}) \right\} .
\end{equation}
If the system is boosted, the wave function becomes
\begin{equation}\label{14}
\psi _{\eta }(z,t) = \left({1 \over \pi }\right)^{1/2}
\exp \left\{-{1\over 2}\left(e^{-2\eta }u^{2} +
e^{2\eta }v^{2}\right)\right\} .
\end{equation}
We note here that the transition from Eq.(\ref{13}) to Eq.(\ref{14}) is a
squeeze transformation.  The wave function of Eq.(\ref{13}) is distributed
within a circular region in the $u v$ plane, and thus in the $z t$ plane.
On the other hand, the wave function of Eq.(\ref{14}) is distributed in an
elliptic region.  This ellipse is a ``squeezed'' circle with the same area
as the circle on the $zt$ plane.

\section{Feynman's Parton Picture}\label{parton}

It is safe to believe that hadrons are quantum bound states of quarks having
localized probability distribution.  As in all bound-state cases, this
localization condition is responsible for the existence of discrete mass
spectra.  The most convincing evidence for this bound-state picture is the
hadronic mass spectra which are observed in high-energy
laboratories~\cite{fkr71,knp86}.
However, this picture of bound states is applicable only to observers
in the Lorentz frame in which the hadron is at rest.  How would the
hadrons appear to observers in other Lorentz frames?  More specifically,
can we use the picture of Lorentz-squeezed hadrons discussed in
Sec.~\ref{covham}.

The radius of the proton is $10^{-5}$ of that of the hydrogen atom.
Therefore, it is not unnatural to assume that the proton has a point
charge in atomic physics.  However, while carrying out experiments on
electron scattering from proton targets, Hofstadter in 1955 observed
that the proton charge is spread out~\cite{hofsta55}.

In this experiment, an electron emits a virtual photon, which
then interacts with the proton.  If the proton consists of quarks
distributed within a finite space-time region, the virtual photon will
interact with quarks which carry fractional charges.  The scattering
amplitude will depend on the way in which quarks are distributed within the
proton.  The portion of the scattering amplitude which describes the
interaction between the virtual photon and the proton is called the form
factor.

Although there have been many attempts to explain this phenomenon within the
framework of quantum field theory, it is quite natural to expect that the
wave function in the quark model will describe the charge distribution.  In
high-energy experiments, we are dealing with the situation in which the
momentum transfer in the scattering process is large.  Indeed, the
Lorentz-squeezed wave functions lead to the correct behavior of the hadronic
form factor for large values of the momentum transfer~\cite{fuji70}.

While the form factor is the quantity which can be extracted from the
elastic scattering, it is important to realize that in high-energy
processes, many particles are produced in the final state.  They are called
inelastic processes.  While the elastic process is described by the total
energy and momentum transfer in the center-of-mass coordinate system, there
is, in addition, the energy transfer in inelastic scattering.  Therefore, we
would expect that the scattering cross section would depend on the energy,
momentum transfer, and energy transfer.  However, one prominent feature in
inelastic scattering is that the cross section remains nearly constant for a
fixed value of the momentum-transfer/energy-transfer ratio.  This phenomenon
is called ``scaling''~\cite{bj69}.

In order to explain the scaling behavior in inelastic scattering, Feynman in
1969 observed that a fast-moving hadron can be regarded as a collection of
many ``partons'' whose properties do not appear to be identical to those of
quarks~\cite{fey69}.  For example, the number of quarks inside a static
proton is three, while the number of partons in a rapidly moving proton
appears to be infinite.  The question then is how the proton looking like a
bound state of quarks to one observer can appear different to an observer in
a different Lorentz frame?  Feynman made the following systematic
observations.

    a). The picture is valid only for hadrons moving with velocity close
       to that of light.

    b). The interaction time between the quarks becomes dilated, and
        partons\\
\hspace{20mm} behave as free independent particles.

    c). The momentum distribution of partons becomes widespread as the
       hadron\\
\hspace{20mm} moves fast.

    d). The number of partons seems to be infinite or much larger than that
       of quarks.

\noindent Because the hadron is believed to be a bound state of two or three
quarks, each of the above phenomena appears as a paradox, particularly b) and
c) together.  We would like to resolve this paradox using the covariant
harmonic oscillator formalism.

For this purpose, we need a momentum-energy wave function.  If the quarks
have the four-momenta $p_{a}$ and $p_{b}$, we can construct two independent
four-momentum variables~\cite{fkr71}
\begin{equation}
P = p_{a} + p_{b} , \qquad q = \sqrt{2}(p_{a} - p_{b}) .
\end{equation}
The four-momentum $P$ is the total four-momentum and is thus the hadronic
four-momentum.  $q$ measures the four-momentum separation between the quarks.

We expect to get the momentum-energy wave function by taking the Fourier
transformation of Eq.(\ref{14}):
\begin{equation}\label{fourier}
\phi_{\eta }(q_{z},q_{0}) = \left({1 \over 2\pi }\right)
\int \psi_{\eta}(z, t) \exp{\left\{-i(q_{z}z - q_{0}t)\right\}} dx dt .
\end{equation}
Let us now define the momentum-energy variables in the light-cone coordinate
system as
\begin{equation}\label{conju}
q_{u} = (q_{0} - q_{z})/\sqrt{2} ,  \qquad
q_{v} = (q_{0} + q_{z})/\sqrt{2} .
\end{equation}
In terms of these variables, the Fourier transformation of
Eq.(\ref{fourier}) can be written as
\begin{equation}\label{fourier2}
\phi_{\eta }(q_{z},q_{0}) = \left({1 \over 2\pi }\right)
\int \psi_{\eta}(z, t) \exp{\left\{-i(q_{u} u + q_{v} v)\right\}} du dv .
\end{equation}
The resulting momentum-energy wave function is
\begin{equation}\label{phi}
\phi_{\eta }(q_{z},q_{0}) = \left({1 \over \pi }\right)^{1/2}
\exp\left\{-{1\over 2}\left(e^{-2\eta}q_{u}^{2} +
e^{2\eta}q_{v}^{2}\right)\right\} .
\end{equation}
Because we are using here the harmonic oscillator, the mathematical form
of the above momentum-energy wave function is identical to that of the
space-time wave function.  The Lorentz squeeze properties of these wave
functions are also the same.  This aspect of the squeeze has been
exhaustively discussed in the literature~\cite{knp86,kim89}.

When the hadron is at rest with $\eta = 0$, both wave functions behave like
those for the static bound state of quarks.  As $\eta$ increases, the wave
functions become continuously squeezed until they become concentrated along
their respective positive light-cone axes.  Let us look at the z-axis
projection of the space-time wave function.  Indeed, the width of the quark
distribution increases as the hadronic speed approaches that of the speed of
light.  The position of each quark appears widespread to the observer in the
laboratory frame, and the quarks appear like free particles.

The momentum-energy wave function is just like the space-time wave function.
The longitudinal momentum distribution becomes wide-spread as the hadronic
speed approaches the velocity of light.  This is in contradiction with our
expectation from nonrelativistic quantum mechanics that the width of the
momentum distribution is inversely proportional to that of the position wave
function.  Our expectation is that if the quarks are free, they must have
their sharply defined momenta, not a wide-spread distribution.  This apparent
contradiction presents to us the following two fundamental questions:

   a).  If both the spatial and momentum distributions become widespread
      as the hadron moves, and if we insist on Heisenberg's uncertainty
      relation, is Planck's constant dependent on the hadronic velocity?

   b).  Is this apparent contradiction related to another apparent
      contradiction that the number of partons is infinite while there
      are only two or three quarks inside the hadron?

The answer to the first question is ``No'', and that for the second question
is ``Yes''.  Let us answer the first question which is related to the Lorentz
invariance of Planck's constant.  If we take the product of the width of the
longitudinal momentum distribution and that of the spatial distribution, we
end up with the relation
\begin{equation}
<z^{2}><q_{z}^{2}> = (1/4)[\cosh(2\eta)]^{2}  .
\end{equation}
The right-hand side increases as the velocity parameter increases.  This
could lead us to an erroneous conclusion that Planck's constant becomes
dependent on velocity.  This is not correct, because the longitudinal
momentum variable $q_{z}$ is no longer conjugate to the longitudinal
position variable when the hadron moves.

In order to maintain the Lorentz-invariance of the uncertainty product,
we have to work with a conjugate pair of variables whose product does
not depend on the velocity parameter.  Let us go back to Eq.(\ref{conju})
and Eq.(\ref{fourier2}).  It is quite clear that the light-cone variable
$u$ and $v$ are conjugate to $q_{u}$ and $q_{v}$ respectively.  It is
also clear that the distribution along the $q_{u}$ axis shrinks as the
$u$-axis distribution expands.  The exact calculation leads to
\begin{equation}
<u^{2}><q_{u}^{2}> = 1/4 , \qquad  <v^{2}><q_{v}^{2}> = 1/4  .
\end{equation}
Planck's constant is indeed Lorentz-invariant.

Let us next resolve the puzzle of why the number of partons appears to
be infinite while there are only a finite number of quarks inside the
hadron.  As the hadronic speed approaches the speed of light, both the
x and q distributions become concentrated along the positive light-cone
axis.  This means that the quarks also move with velocity very close
to that of light.  Quarks in this case behave like massless particles.

We then know from statistical mechanics that the number of massless
particles is not a conserved quantity.  For instance, in black-body
radiation, free light-like particles have a widespread momentum
distribution.  However, this does not contradict the known principles
of quantum mechanics, because the massless photons can be divided into
infinitely many massless particles with a continuous momentum
distribution.

Likewise, in the parton picture, massless free quarks have a wide-spread
momentum distribution.  They can appear as a distribution of an
infinite number of free particles.  These free massless particles are the
partons.  It is possible to measure this distribution in high-energy
laboratories, and it is also possible to calculate it using the covariant
harmonic oscillator formalism.  We are thus forced to compare these two
results.  Indeed, according to Hussar's calculation~\cite{hussar81},
the Lorentz-boosted oscillator wave function produces a reasonably
accurate parton distribution.

\section{Coherence Problems}\label{cohere}
The most puzzling problem in the parton picture is that partons in
the hadron appear as incoherent particles, while quarks are coherent
when the hadron is at rest.  Does this mean that the coherence is
destroyed by the Lorentz boost?   The answer is NO, and here is the
resolution to this puzzle.

When the hadron is boosted, the hadronic matter becomes squeezed and
becomes concentrated in the elliptic region along the positive
light-cone axis.  The length of the major axis becomes expanded by
$e^{\eta}$, and the minor axis is contracted by $e^{\eta}$.

This means that the interaction time of the quarks among themselves
become dilated.  Because the wave function becomes wide-spread, the
distance between one end of the harmonic oscillator well and the
other end increases.  This effect, first noted by Feynman~\cite{fey69},
is universally observed in high-energy hadronic experiments.  The
period is oscillation is increases like $e^{\eta}$.

On the other hand, the interaction time with
the external signal, since it is moving in the direction opposite to
the direction of the hadron, it travels along the negative light-cone
axis.  If the hadron contracts along the negative light-cone axis, the
interaction time decreases by $e^{-\eta}$.  The ratio of the interaction
time to the oscillator period becomes $e^{-2\eta}$.  The energy of each
proton coming out of the Fermilab accelerator is $900 GeV$.  This leads
the ratio to $10^{-6}$.  This is indeed a small number.  The external
signal is not able to sense the interaction of the quarks among
themselves inside the hadron.

\section*{Concluding Remarks}
The time variable plays many important roles in physics.  Its place in
Einstein's special relativity is well known.  In this paper, we noted
that the time-separation variable together with the spatial separation
can be combined into one covariant world in the quark-parton model of
relativistic hadrons.
The time separation variable plays also plays a pivotal role in the
measurement process in the parton picture where the partons appear like
incoherent entities.  It is shown that the lack of coherence in the
parton picture is perfectly consistent with special relativity.

\end{document}